\documentclass[preprint,sort&compress]{elsarticle}
\usepackage{textcomp}
\usepackage{siunitx}
\usepackage{amsmath}
\usepackage{tabularx}
\usepackage{graphicx}
\usepackage{array}
\usepackage{makecell}
\usepackage{caption}
\usepackage{hyperref}
\usepackage[dvipsnames,table,xcdraw]{xcolor}
\usepackage[normalem]{ulem}
\usepackage{pdfpages}
\definecolor{linkcolor}{rgb}{0.5,0.1,0.1}
\definecolor{urlcolor} {rgb}{0.1,0.1,0.5}
\definecolor{citecolor}{rgb}{0.1,0.5,0.1}
\hypersetup{colorlinks=true,linkcolor=linkcolor,urlcolor=urlcolor,citecolor=citecolor,pdfborder={0 0 0}}
\journal{Journal of Colloid and Interface Science}

\begin{document}
\begin{frontmatter}
\title{Scalable Size- and Shape-Selective Purification of Colloidal Building Blocks via Excluded Volume Interactions}
\author[1,2]{Thomas Kainz}
\author[1,2]{Isobel McSweeney}
\author[1,2]{Andrei-Micea Top}
\author[1,2]{Nicolas Bruder}
\author[3]{David J. Pine}
\author[1,2]{Andrea Dodero\corref{cor1}{}}
\ead{andrea.dodero@unifr.ch}
\author[1,2]{Ullrich Steiner\corref{cor1}{}}
\ead{ullrich.steiner@unifr.ch}
\cortext[cor1]{Corresponding authors}
\affiliation[1]{Adolphe Merkle Institute, University of Fribourg, Chemin des Verdiers 4, 1700 Fribourg, Switzerland}
\affiliation[2]{National Competence Center in Bioinspired Materials, University of Fribourg, Chemin des Verdiers 4, 1700 Fribourg, Switzerland}
\affiliation[3]{Department of Physics, New York University, 726 Broadway, New York, NY 10003, USA}

\begin{abstract}
Excluded-volume interactions, arising solely from steric constraints, play a crucial role in determining the structure, dynamics, and phase behaviour of colloidal suspensions. This is particularly important for non-spherical particles, where orientation-dependent effects also become significant. In this study, we employ depletion-driven phase separation to develop a scalable, size-selective method for purifying spherical and non-spherical colloidal clusters that exhibit an interplay of concave and convex surface areas. Phase diagrams of charge-stabilised polystyrene spheres ranging in size from 267 to 1008 nm demonstrate that the mixing-demixing transition occurs across a range of surfactant concentrations rather than at a single threshold. Taking advantage of this transition width enables the purification of a single component from binary mixtures at size ratios as low as 1.6 in a single step. When the same approach is applied to tetrameric colloidal clusters, these are enriched fifteenfold relative to uncoordinated spheres. Importantly, the efficiency of sorting depends not only on the effective size but also on the geometry of the aggregate. For instance, anisotropic, weakly fused clusters separate more efficiently than spherical aggregates because their concave surface curvature is reduced compared to unfused clusters. These findings establish excluded-volume-driven sorting as a practical and scalable route for purifying colloidal building blocks for hierarchical assembly.
\end{abstract}
\end{frontmatter}

\section{Introduction}
\noindent Interactions between particles in colloidal suspensions determine their structure, dynamics, and phase behaviour \cite{poon2004, C5SM02038G}. Among others, excluded volume interactions are a defining feature of colloidal suspensions, being entropic in origin and arising from purely steric, repulsive forces \cite{frenkel1988, C8SM02048E}. For spherical colloids, excluded volume interactions are well understood and underpin classical theories of colloidal phase behaviour \cite{verwey1948, Poon2002Phase}. However, many natural and synthetic colloids are non-spherical, ranging from rods, plates, and polyhedra to more complex anisotropic geometries \cite{Glotzer2007Anisotropy}. In such systems, excluded volume effects depend not only on concentration but also sensitively on orientation. This leads to entropic forces that drive a range of phenomena, such as liquid-crystalline order, directional self-assembly, and anisotropic mechanical responses \cite{onsager1949,doi1986, AVENDANO201762}. Understanding these orientation-dependent excluded volume interactions between non-spherical colloidal particles is therefore essential for developing predictive theories of anisotropic soft matter and for guiding the design of advanced materials with tunable structural and functional properties \cite{Glotzer2007Anisotropy,frenkel1999,doi:10.1126/science.1253751}.

Indeed, beyond governing simple equilibrium phase behaviour, excluded volume can be harnessed in various ways to manipulate colloidal suspensions. In particular, excluded volume interactions can drive entropic phase separation, thereby enabling the efficient separation of colloidal objects of differing sizes \cite{AsakuraOosawa54, Lekkerkerker2011Colloids, Lekkerkerker2024Depletion}. When combined with gravitational sedimentation or creaming, this approach can also be used to accelerate colloidal crystallisation into face-centred cubic (FCC) and hexagonal close-packed (HCP) opals \cite{SANDERS:1964aa, PuseyNature1986}. 
While these early approaches addressed the phase equilibria of different globular objects (typically colloids and dilute polymer coils), depletion interactions can also be employed to separate colloids of the same type based on size 
\cite{AsakuraOosawa54, AsakuraOosawa58, Israelachvili2011}. Thirty-five years ago, Bibette demonstrated that adding surfactant micelles to a polydisperse colloidal mixture induces size-selective aggregation, separating the mixture into two fractions containing large and small particle populations \cite{Bibette1991, SteinerEmulsions1995}.
\begin{figure*}[tbp]
  \includegraphics[width=\linewidth]{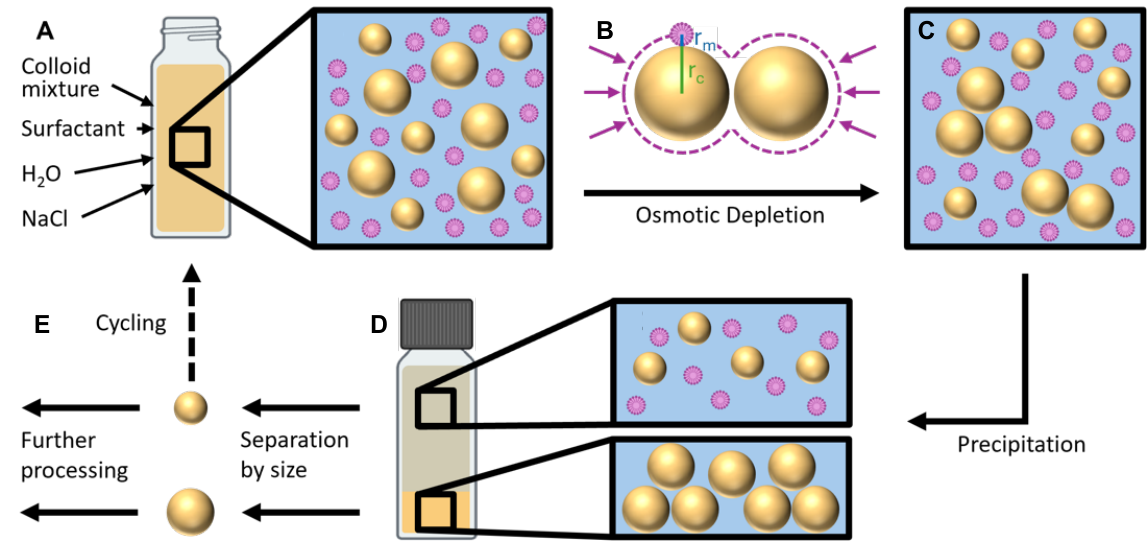}
  \caption{Schematic of the colloidal sorting mechanism: (A) A surfactant is added to a binary colloidal (yellow) suspension, forming micelles (pink). (B) The distance between the centres of micelles and the colloid surface defines the excluded volume (dashed line). When two colloidal spheres approach each other, their excluded volumes overlap, expelling the micelles from the gap between them. The uncompensated osmotic pressure (purple arrows) then causes the two spheres to aggregate. (C) For a given number of micelles, which is determined by the surfactant concentration, only larger spheres aggregate. (D) This causes their sedimentation, leaving only the smaller colloidal spheres in suspension and enabling the size-sorting of colloids (E).}
  \label{fig:overview}
\end{figure*}

This three-particle approach is known as the `Bibette method' and is illustrated schematically in Figure \ref{fig:overview}. When the surfactant concentration exceeds the critical micelle concentration (CMC) \cite{Rosen2012Surfactants}, nanometre-sized micelles form in the colloidal suspension (purple spheres in Fig.\,\ref{fig:overview}). Hard-core repulsion prevents micelles from approaching colloids closer than a micelle radius, thereby defining a `depletion volume' \cite{Vrij1976Depletion, XING201554, BRISCOE201546}, represented by a purple dashed line in Fig.\,\ref{fig:overview}B. When two colloids approach each other, their depletion volumes overlap, and the small micelles are therefore excluded from the gap. This results in an uncompensated osmotic pressure exerted by the micelles on the two colloids (purple arrows in Fig.\,\ref{fig:overview}B), which causes the larger colloids to aggregate and ultimately to flocculate.  This classical picture, originally developed for micrometre-scale colloids, has recently been shown to hold quantitatively down to the nanoscale, with deviations only appearing once the depletant size becomes comparable to that of the colloid \cite{Ofosu2025NanocrystalDepletion}.

An equivalent thermodynamic description considers the entropy of the suspension. For spherical objects (i.e., colloids and micelles), the entropy is given by the sum of all their positional conformations within the sample volume. When colloids are packed, their excluded volumes overlap, which increases the free volume available to the micelles \cite{Lekkerkerker2011Colloids}, thereby increasing their entropy. When micelles are far more numerous than colloids, this gain in micelle entropy outweighs the loss in colloid entropy, driving phase separation into a colloid- and micelle-rich phase. 

The Bibette method, therefore, relies on the interplay of two parameters: (i) the volume `gained' when two colloidal spheres approach each other, which depends on the radii, and (ii) the ensuing entropy gain of the micelles, which scales with the number of micelles and thus the surfactant concentration. For the binary colloidal mixture schematically shown in Fig.\,\ref{fig:overview}E, it is possible to identify a surfactant concentration at which the larger spheres flocculate. In contrast, the smaller ones remain in suspension (Fig.\,\ref{fig:overview}C, D), enabling scalable size separation of large and small colloidal spheres (Fig.\,\ref{fig:overview}E) \cite{Bibette1991}. This approach can be repeated to purify monodisperse size fractions from an initial polydisperse suspension \cite{SteinerEmulsions1995}.

Although initial models of depletion interactions involving spherical colloids were extended to include non-spherical objects (see above, \cite{Glotzer2007Anisotropy}), the recent design of colloidal clusters presents new challenges. A notable example is the formation of a colloidal cluster comprising four `hard' spheres held together by a central `soft' oil droplet \cite{Sacanna2011Nature, PineNature2020} (see Fig.\,S1).
The synthesis, colloidal-molecule assembly, and design principles of such patchy colloidal clusters have recently been reviewed in detail \cite{Kim2023PatchyColloidalClusters}. The resulting tetrahedron-shaped colloidal object features an interplay of convex and concave surface curvatures. Qualitatively, depletion interactions mediated by small spheres (e.g., surfactant micelles) predominantly occur between convex surfaces, whereas concave surfaces do not contribute to this effect. For example, this has been demonstrated in lithographically manufactured platelets, where an increase of surface roughness resulted in weaker micelle-mediated depletion interactions \cite{PhysRevLett.99.268301}.

While this work was originally developed for colloidal sorting, it differs from earlier approaches in that it considers colloidal clusters that exhibit an interplay of convex and concave surface areas.

Here, we analyse mixtures of colloidal aggregates using the Bibette method. This method was originally developed for colloidal sorting and is now a well-established process \cite{Lekkerkerker2024Depletion,Yang2026FractionalCrystallization}. Our work differs from earlier approaches in that it considers colloidal clusters that exhibit an interplay of convex and concave surface areas. Our focus is on separating (i) colloidal clusters from uncoordinated polystyrene (PS) spheres and (ii) tetramers from other cluster types. First, we determine the conditions required to achieve high-level size purification in a single experimental step. Secondly, we use this method to separate colloidal aggregates, such as tetramers and trimers, from isolated PS spheres and from each other. This demonstrates that separation efficiency depends heavily on aggregate shape. Finally, we provide an overview of how depletion interactions depend on the interplay of convex and concave surface curvatures of various colloidal aggregates.

\section{Results}
\noindent Sorting colloidal spheres by size via depletion interactions requires knowing the surfactant concentration at which a given sphere size starts to flocculate. This dependence is captured by a colloidal stability phase diagram that maps the onset of depletion-induced phase separation as a function of surfactant concentration and colloid diameter \cite{Poon2002Phase}.

\begin{figure*}[tbp]
  \includegraphics[width=\linewidth]{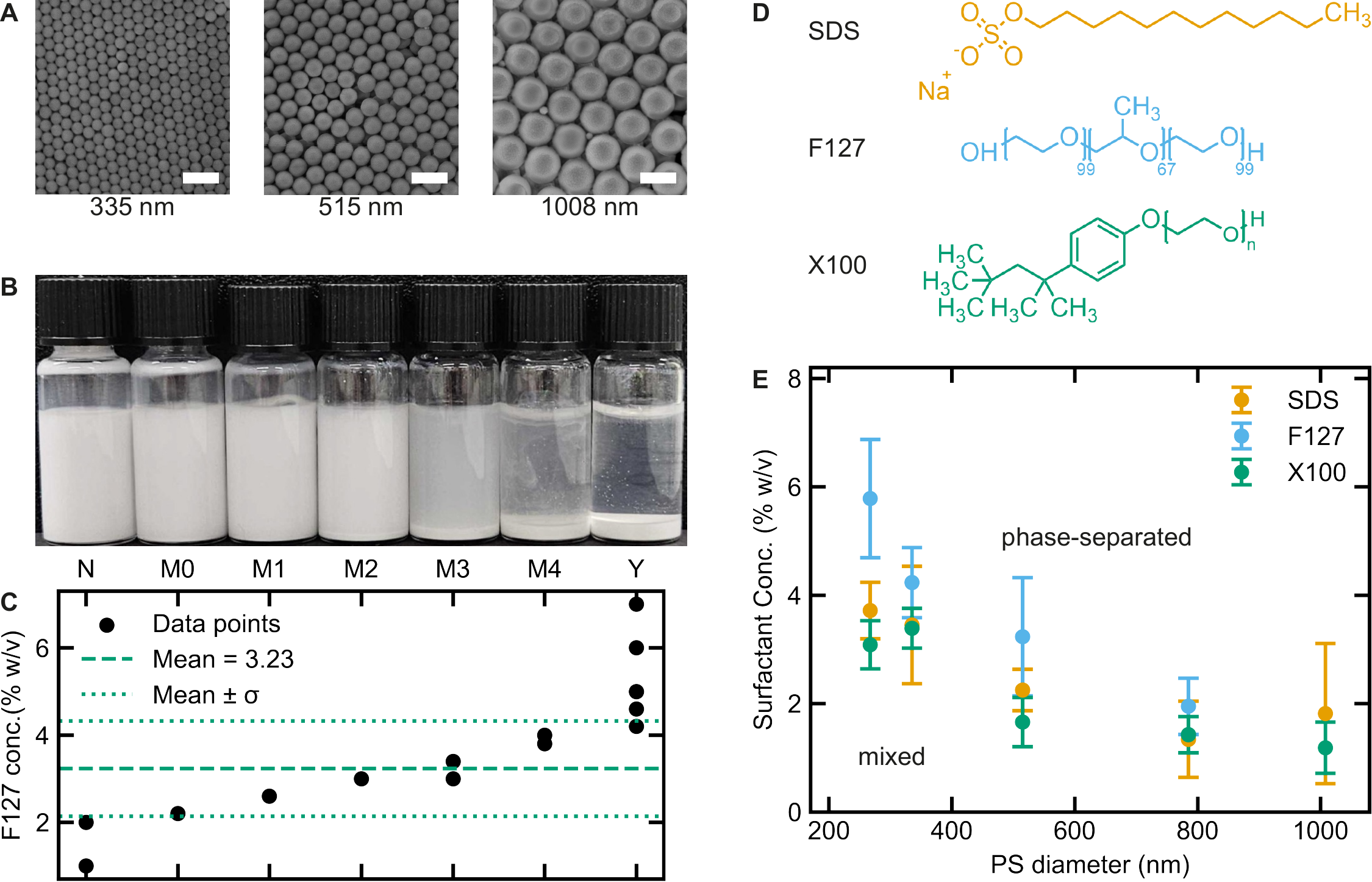}
  \caption{(A) Scanning electron microscope images of three representative polystyrene colloidal sphere samples with low polydispersity. The $\SI{335} {nm}$ and $\SI{515} {nm}$ batches were synthesised using surfactant-free emulsion polymerisation. The $\SI{1008} {nm}$ batch was prepared using a seeded growth method. All scale bars are $\SI{1}{\um}$. (B) Digital photograph showing different levels of depletion present in sample vials containing 515\,nm PS batches. The degree of depletion increases from none on the left to full on the right, with five intermediate levels (see Experimental Section). The shown vials are the same as those analysed in (C). Stronger depletion corresponds to a higher depletant concentration, due to a higher concentration of the micelle-forming surfactant in the mixture. The mean concentration and width of the depletion transition are then calculated. (D) Micelle-forming surfactants investigated in this work: Sodium dodecyl sulfate (SDS), Pluronic F127, and Triton X-100. (E) Phase diagrams for the three surfactants (colour code as in D) with polystyrene batches ranging in diameter from $\SI{267} {nm}$ to $\SI{1008} {nm}$. The width of the transition from the mixed to the demixed phase is indicated by the error bars.
  }
  \label{fig:depletion}
\end{figure*}

To construct these diagrams, five batches of monodisperse polystyrene colloidal suspensions ranging in size from 267 to 1008 nm were prepared, with some examples shown in Fig.\,\ref{fig:depletion}A. These suspensions were tested using three micelle-forming surfactants, namely Pluronic F127, sodium dodecyl sulfate, and Triton X-100 (Fig.\,\ref{fig:depletion}D, see Experimental Section). The colloid concentration was adjusted to 1\,--\,2.5\% (w/v), with the ionic strength fixed at 100\,mM using NaCl. Within this range, the colloid volume fraction in suspension has only a weak effect on the observed depletion threshold. Indeed, provided the number density of surfactant micelles exceeds that of the colloidal spheres by several orders of magnitude, excluded volume interactions are mainly dominated by surfactant concentration. However, because PS colloids are charge-stabilised in suspension, Coulomb repulsion counteracts depletion attractions. Therefore, the Debye screening length must remain smaller than the micellar diameters. At 100 mM NaCl, the Debye screening length is below 1 nm \cite{Hunter2001Zeta}, ensuring that screening is sufficiently strong under the experimental conditions tested.

Figure \ref{fig:depletion}B shows digital images of a representative series of PS colloidal suspensions (diameter of 515\,nm) as the Pluronic F127 concentration is increased, which is shown in Fig.\,\ref{fig:depletion}C. With increasing surfactant concentration, depletion-driven aggregation becomes apparent, as evidenced by the formation of a sediment accompanied by a progressive clarification of the supernatant. Here, two limiting cases can be identified: (i) a fully mixed suspension at a surfactant concentration just below the onset of sedimentation (Fig.\,\ref{fig:depletion}B, left), and (ii) a surfactant concentration above which the supernatant phase remains transparent and unchanged (Fig.\,\ref{fig:depletion}B, right). These two limiting cases define a transition region between fully mixed and fully phase-separated suspensions, as indicated by the dotted lines in Fig. \ref{fig:depletion}C. Note that phase formation is fully reversible, i.e., the sediment can be resuspended by agitating the vials (see Fig.\,S2).

We then show five intermediate stages of sediment formation, each accompanied by a reduction in the cloudiness of the supernatant (Fig.\,\ref{fig:depletion}B, from left to right). From this discretised transition curve, the mean and weighted standard deviation of the phase transition were also calculated (see the Experimental Section).

We emphasise that the aim of our excluded-volume framework is statistical sorting rather than perfect demixing. As the system approaches complete phase separation, we anticipate that contributions to thermodynamics will arise from rotational entropy and from interactions between local exclusion and global available volume. We also anticipate that these contributions will become increasingly significant. Ultimately, these contributions will limit the achievable purity (see Fig.\,S2). Resolving this quantitatively would require an experimental setup dedicated to this purpose, as well as simulation theory capable of explicitly treating concave geometries -- both of which lie beyond the scope of this work.

Repeating this procedure for all colloid sizes and surfactants yields the phase diagram shown in Fig.\,\ref{fig:depletion}E. For all three surfactants, the phase boundary shifts systematically with colloid diameter, enabling identification of surfactant concentrations at which larger spheres flocculate. In comparison, smaller spheres remain dispersed, a process forming the basis of the Bibette method \cite{10.1063/5.0173340}. Notably, the Pluronic F127 phase boundary lies at a slightly higher surfactant concentration, consistent with its significantly larger micellar size (i.e., 20\,nm \cite{gentile2010thermogelation}) compared to SDS (i.e., 6\,nm \cite {mirgorod2019structure}) and Triton X-100 (i.e., 10\,nm \cite{stubivcar1989size}). Such a result can be explained by the fact that, at a given micelle number density, smaller micelles increase the effective excluded-volume interaction between the colloids. Additionally, larger micelles incorporate more surfactant molecules than smaller ones. Together, these effects explain the requirement for higher F127 concentrations.

An observation in Fig. \ref{fig:depletion}B reveals that the phase separation is not sharp but occurs over a certain surfactant concentration window, as captured by the large error bars in Fig.\,\ref{fig:depletion}E. The relatively wide transition region complicates particle-size separation because partially depleted conditions lead to cross-contamination between the supernatant and sediment, thereby reducing the achievable purity in a single step and often necessitating iterative fractionation. This issue becomes particularly relevant for colloidal clusters, where the effective size ratio between the large ($l$) and small ($s$) colloids in suspension approaches $r_\mathrm{l}/r_\mathrm{s}$ of 2 and below, and the transition windows of the two populations can overlap substantially.

\begin{figure} [tbp]
  \includegraphics[width=\linewidth]{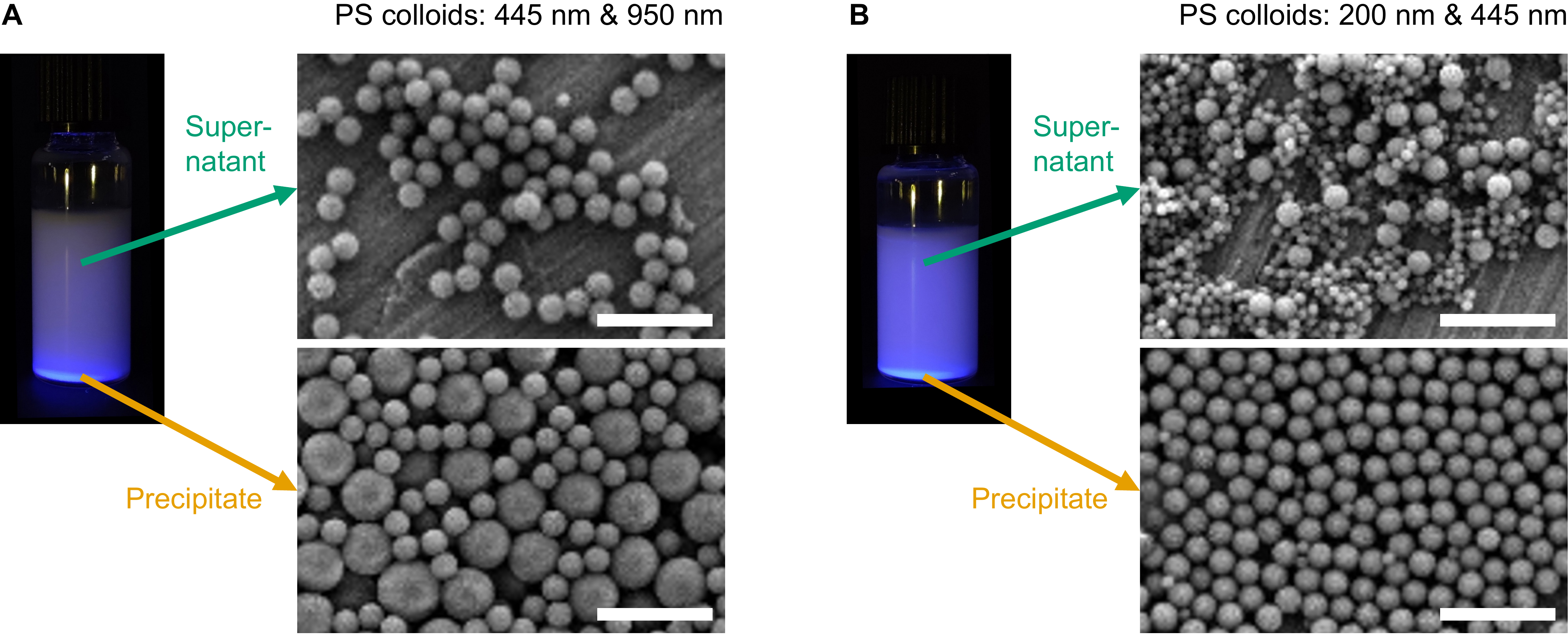}
  \caption{Separation of binary PS colloidal mixtures with a size ratio of $r_\mathrm{l}/r_\mathrm{s}$\,=\,2.2.  
  (A) Blend of 445\,nm and 950\,nm colloids, with a total colloid concentration of 1.2\% w/v. Adding 2.8\% w/v Triton enabled the larger colloids to precipitate completely while keeping the smaller ones in suspension. This resulted in a nearly pure supernatant and a mixed precipitate. (B) Blend of 200\,nm and 445\,nm colloids with a total colloid concentration of 1\% w/v. A 5.5\% w/v Triton concentration was chosen to lie far above the transition region of the larger colloids, yet still within the mixed region of the smaller ones. This produced a pure precipitate and a mixed supernatant. All scale bars are 10\,\textmu m.
}
  \label{fig:sorting}
\end{figure}

Figure \ref{fig:sorting} illustrates depletion-driven phase separation in binary colloidal blends with an $r_\mathrm{l}/r_\mathrm{s}$ ratio of around 2.2 using Triton X-100 as surfactant. Two relevant regimes can be distinguished. In the first one, the surfactant concentration is chosen such that the large spheres are well within the demixing regime, whereas the small spheres are close to (or within) the transition window. Under these conditions, the majority of the large spheres sediment efficiently. In contrast, only a fraction of the small spheres partitions into the sediment, as determined by the phase diagram in Fig.\,\ref{fig:depletion}. This yields a nearly pure supernatant containing small spheres (Fig.\,\ref{fig:sorting}A). Conversely, in the second regime, the surfactant concentration is selected such that the small spheres remain fully mixed, while the large spheres demix strongly. In this case, the sediment contains only large spheres, whereas the supernatant remains mixed due to incomplete removal of large spheres (Fig.\,\ref{fig:sorting}B). 

Despite the thermodynamic limitations described above, Fig.\,\ref{fig:sorting} demonstrates that a single-size population can be purified from a binary blend in a single depletion step, and that either the small or large spheres can be targeted by selecting the surfactant concentration relative to the two transition windows. It is worth noting that the yield can be further increased by repeating the same separation protocol. For a size ratio $r_\mathrm{l}/r_\mathrm{s}$ of around 2.1 an efficient sorting is achieved after three iterations, as shown for each cycle in Fig. S2. A very low loss of the large target particles is achieved. For further repetitions, the thermodynamic demixing limits manifest visibly, reducing the efficiency of sorting. To this end, either the supernatant can be retrieved, or the precipitate can be harvested and redispersed. 
\begin{figure} [tbp]
  \includegraphics[width=\linewidth]{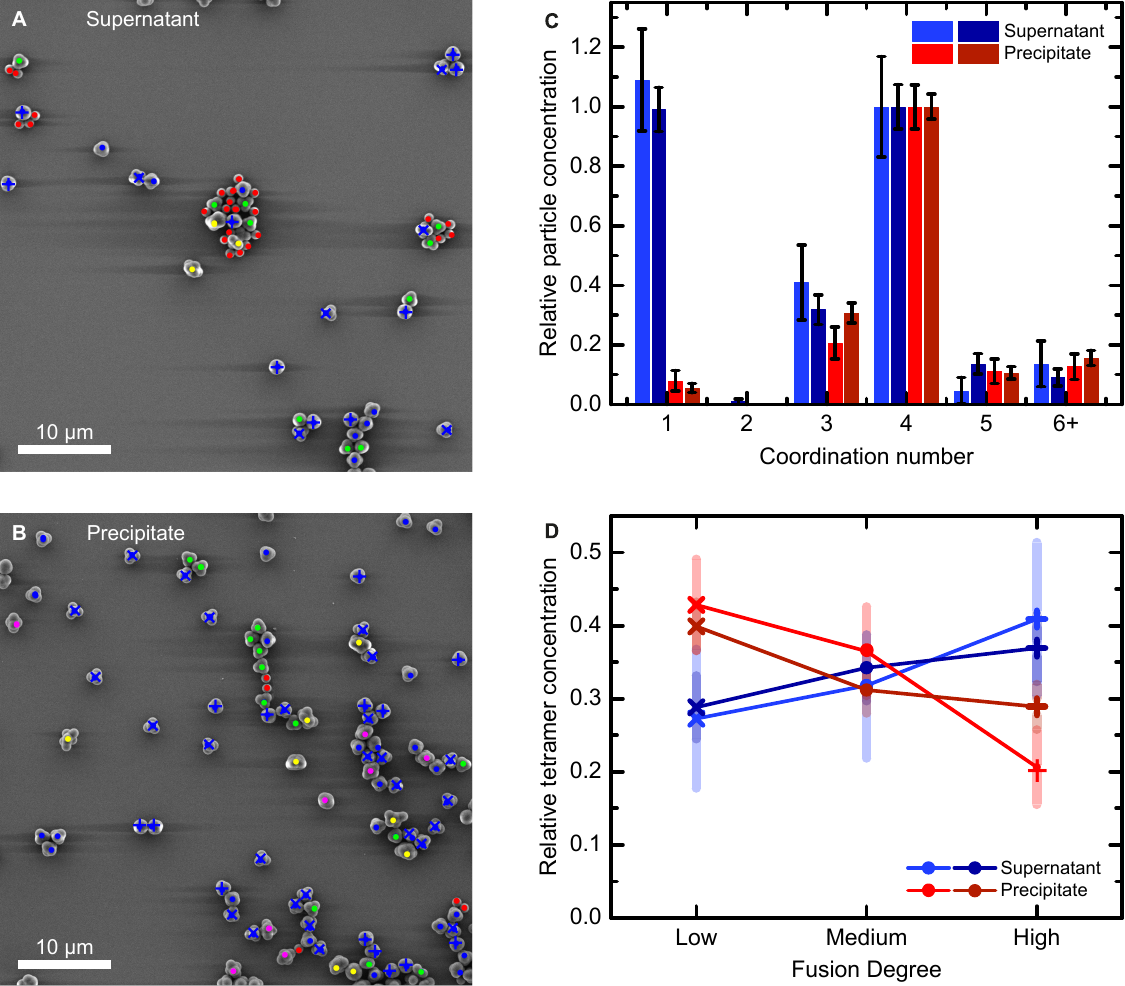}
  \caption{Sorting of colloidal aggregates. Colloidal aggregates consisting of 1008\,nm PS colloids and 478\,nm TPM droplets were prepared according to the method described in the Experimental Section and in the SI. The resulting 1.01\% w/v suspension was supplemented with $0.8\%$ w/v Triton X-100 and 101\,mM NaCl. Over 24\,h, the blend separated into supernatant and sediment, both of which were analysed by SEM ((A), (B), respectively). The clusters were labelled manually: red, cyan, green, blue, and magenta for aggregate numbers 1-5, respectively. Hexamers and higher-order aggregates are marked in yellow. The sediment (B) contained visibly more tetramers and fewer uncoordinated spheres than the supernatant (A). The different aggregates were counted, and their relative numbers were plotted (C; supernatant: light blue; precipitate: light red). SEM images at five-times lower magnification were also analysed (see SI; supernatant: dark blue; precipitate: dark red). The tetramer-to-sphere ratio in the sediment was 15 times higher than in the supernatant (1:1). No distinct sorting beyond the statistical error was found for coordination numbers of three or more.
  The tetramers were further subdivided according to their level of deformation (see text).
   The relative concentrations of the different fusion degrees are normalised by the total number of tetramers (D).
  In contrast to strongly deformed, more spherical aggregates, less-deformed tetramers exhibiting four visible lobes were more prevalent in the sediment than in the supernatant. This indicates that sorting performance decreases as deformation increases.
  \label{fig:Tsort}}
\end{figure}

Having established the phase diagram and the sorting behaviour of binary sphere mixtures, we now turn our focus to the sorting of colloidal aggregates \cite{https://doi.org/10.1002/anie.201001451}. To this end, colloidal clusters were prepared as illustrated in Fig.\,S1 \cite{Sacanna2013Shaping}. The preparation method is described in detail in the Experimental Section. In brief, a large excess of charge-stabilised 1008 nm polystyrene spheres was mixed with 478\,nm TPM droplets. Typically, four PS spheres assemble around each TPM droplet to form a tetramer, whereas smaller populations of other coordination number aggregates (i.e., trimers, pentamers, etc.) are also formed. The resulting mixture, therefore, consists predominantly of uncoordinated PS spheres and tetramers, with minor fractions of other aggregates. To enable subsequent lattice assembly, which is beyond the scope of this study, partially fused clusters are required \cite{PineNature2020,Neophytou2021ColloidalDiamond, Baran2023PursuingColloidalDiamonds}. This can be achieved by adding tetrahydrofuran (THF) to the colloidal suspension, leading to a swelling and softening of the PS spheres \cite{Chen2011Fusion}. This promotes sphere fusion as the liquid TPM droplet is extruded towards the particle surface, forming patches. After sufficient exposure to THF, larger PS spheres are formed with tetrahedrally arranged TPM patches. However, if THF is removed before full fusion occurs, partially deformed tetramers with residual lobes are obtained. Because the fusion is highly sensitive to THF exposure and mixing, imperfect control can yield a distribution of fusion states. 

In this context, we address two questions: (i) whether colloidal clusters can be separated from uncoordinated PS spheres using depletion-driven phase separation, and (ii) whether clusters differing in coordination number or degree of fusion (i.e., size and shape) can be sorted. For the phase separation experiment, 0.8\% w/v Triton X-100 was added to a 1.01\% w/v cluster suspension, and the ionic strength was adjusted to 101\,mM using NaCl. After 24\,h, the suspension had separated into a sediment and a supernatant, both of which were examined using SEM. Figure \ref{fig:Tsort}A, B show SEM images of the supernatant and the sediment, respectively. Uncoordinated PS spheres and different clusters were identified and manually colour-coded. A cursory examination of the two images reveals the following: (i) a significant difference in the ratio of tetramers (blue) to uncoordinated PS spheres (red), and (ii) the presence of differently fused tetramers. Figure \ref{fig:Tsort}C quantifies the efficiency with which the tetramers were sorted from the uncoordinated spheres, normalised to the total number of particles counted in two sample areas imaged at different magnifications (Figure \ref{fig:Tsort}A, B, Fig.\,S4, Table S1). While the supernatant contains approximately equal numbers of tetramers and PS spheres, the sediment exhibits a ratio of approximately 15:1, indicating the efficiency of the separation process. A careful analysis of the ratios of the different aggregates shows that they remain unchanged within statistical error. Therefore, although it is possible to separate colloidal clusters from uncoordinated PS spheres efficiently, separating colloidal clusters with different coordination numbers is substantially more challenging under these conditions. The analysis parameters are summarised in Table S2. 

Notably, the presence of differently fused clusters indicates that the fusion process was uncontrolled during the sample-to-THF transfer. While this would be undesirable in other experimental settings, it allows us to analyse our data more thoroughly. To better probe shape effects and by focusing exclusively on the tetramers in the supernatant and sediment, we classified them into three fusion categories based on their compression ratio $\eta_\mathrm{c}$: (i) `low-deformation’, where the four colloids extend distinctly from their TPM core and the lobes remain clearly visible, (ii) `high-deformation', where the cluster has fused completely to form a nearly spherical shape, and (iii) `medium-deformation’ aiming at $\eta_\mathrm{c} \!\in \![0.6;0.8]$, the diamond crystallisation range as shown in \cite{PineNature2020}.
A detailed use of the size ratio $\eta_\mathrm{s}$ and the compression ratio $\eta_\mathrm{c} = d_\mathrm{cc}/(2a)$ as the distance between the centres of the spherical lobes divided by twice their radius is shown in Fig.\,S3. The size ratio $\eta_\mathrm{s}\!=\!b/a$, as the distance between the centres of the spherical lobes divided by twice their radius, is used as an additional measure, as it peaks for `medium-deformation’. 
Figure \ref{fig:Tsort}D shows the relative frequency of each class, normalised by the total number of tetramers found in the SEM images. As there are relatively few tetramers in Fig.\,\ref{fig:Tsort}A, B, we also analysed SEM images of the same sample taken at a lower magnification (Fig.\,S4). The analysis parameters are summarised in Table S3. 

The sediment contains significantly more low-deformation (tetrahedral) tet\-ra\-mers, whereas the supernatant contains a higher fraction of highly deformed (i.e., more spherical) tetramers. This demonstrates that sorting is most effective for unfused colloidal clusters with distinct lobes extending from the core (i.e., more anisotropic geometry). Notably, it is possible to separate fully fused (i.e., spherical) tetramers from uncoordinated polystyrene spheres with excellent selectivity by carefully selecting the experimental parameters, as shown in Fig.\,S5.

\section{Discussion}
\noindent Our study revisits the size-based sorting of colloidal mixtures via depletion, a concept first proposed approximately 30 years ago \cite{Bibette1991}. While phase coexistence in depletion-controlled colloidal mixtures has been studied before, our work extends prior work in three key ways. Firstly, we map the mixing-demixing boundary systematically across particle size and surfactant type. Secondly, we quantify the width of the transition region between the mixed and demixed states in binary colloidal blends rather than treating the boundary as a single threshold. Thirdly, we apply the approach to non-spherical colloidal clusters, where geometry can influence both sorting efficiency and kinetics.  Complementary strategies that instead exploit differences in surface chemistry to sort mixed colloidal suspensions have also been reported recently \cite{Tong2026SurfaceDictatedSorting}, underscoring depletion as one of several emerging, scalable routes towards general-purpose colloidal sorting. Fluorescence-activated cell sorting has recently been demonstrated as an alternative, higher-throughput route for purifying colloidal clusters by size, shape, and composition \cite{vanKesteren2024FACS}.

Previous studies of excluded-volume-driven phase separation in colloidal suspensions and emulsions typically report the existence of two coexisting phases above a critical depletant concentration. The results shown in Fig.\,\ref{fig:depletion}B, C demonstrate that the onset of the phase boundary is not sharp. Instead, we observe a range of surfactant concentrations (equivalent to micelle concentrations) over which phase separation occurs. This transition width is central to practical sorting because it determines the extent of cross-contamination between supernatant and sediment under partially depleted conditions. 

The existence of such a transition region may have thermodynamic and/or kinetic origins. For sufficiently stabilised colloidal suspensions, the osmotic pressure effect illustrated in Fig.\,\ref{fig:overview}C is reversible and subject to concentration fluctuations in the mixture. Therefore, qualitatively, the ratio of aggregated to non-aggregated volume should depend on the overall osmotic pressure, i.e.,\ the surfactant concentration. Conversely, for marginally stabilised suspensions (e.g.,\ short Debye lengths in the case of charge stabilisation), colloidal aggregation may be irreversible.

Changing the micelle concentration also alters the driving osmotic pressure responsible for colloidal aggregation, thereby influencing the kinetics of the phase-separation process. This implies that mixtures with lower surfactant concentrations may require longer times to reach their final partitioning. However, quantifying the kinetics of phase separation is challenging because uncoordinated colloids sediment under gravity. Experimental strategies, such as density-matching the suspension or very slow sample rotation, could help decouple depletion-driven demixing from sedimentation and enable a more direct kinetic analysis of phase formation.

Notably, by carefully tuning the surfactant concentration within the transition window of one population but outside that of the other, it is possible to purify one colloidal population while leaving the other unpurified. This is a direct consequence of the width of the phase transition region described above. Validating this scenario requires quantitative composition measurements, since the two phases can remain similarly turbid and cannot be reliably distinguished by eye. Accordingly, we fluorescently labelled a single species \cite{https://doi.org/10.1002/pi.4842} and combined fluorescence readout with SEM analysis to quantify the compositions of the supernatant and precipitate.

We now assess the feasibility of the primary objective of this study: separating coordinated colloidal clusters from uncoordinated polystyrene spheres and, ultimately, sorting clusters by coordination number. In a crude approximation, clusters can be represented as effective spheres. For instance, a tetramer has approximately four times the volume of individual spheres, with a diameter $\sqrt[3]{4}\approx 1.6$ times greater. This size ratio is smaller than those calculated for Fig.\,\ref{fig:sorting}, with size ratios of 2.13 and 2.23. However, it is nevertheless possible to separate spheres of a size ratio of $\approx1.6$, as demonstrated by fully fusing the colloidal clusters into spheres and showing their sorting from uncoordinated spheres in Fig.\,S5.

\begin{figure}
    \centering
    \includegraphics[width=1.0\linewidth]{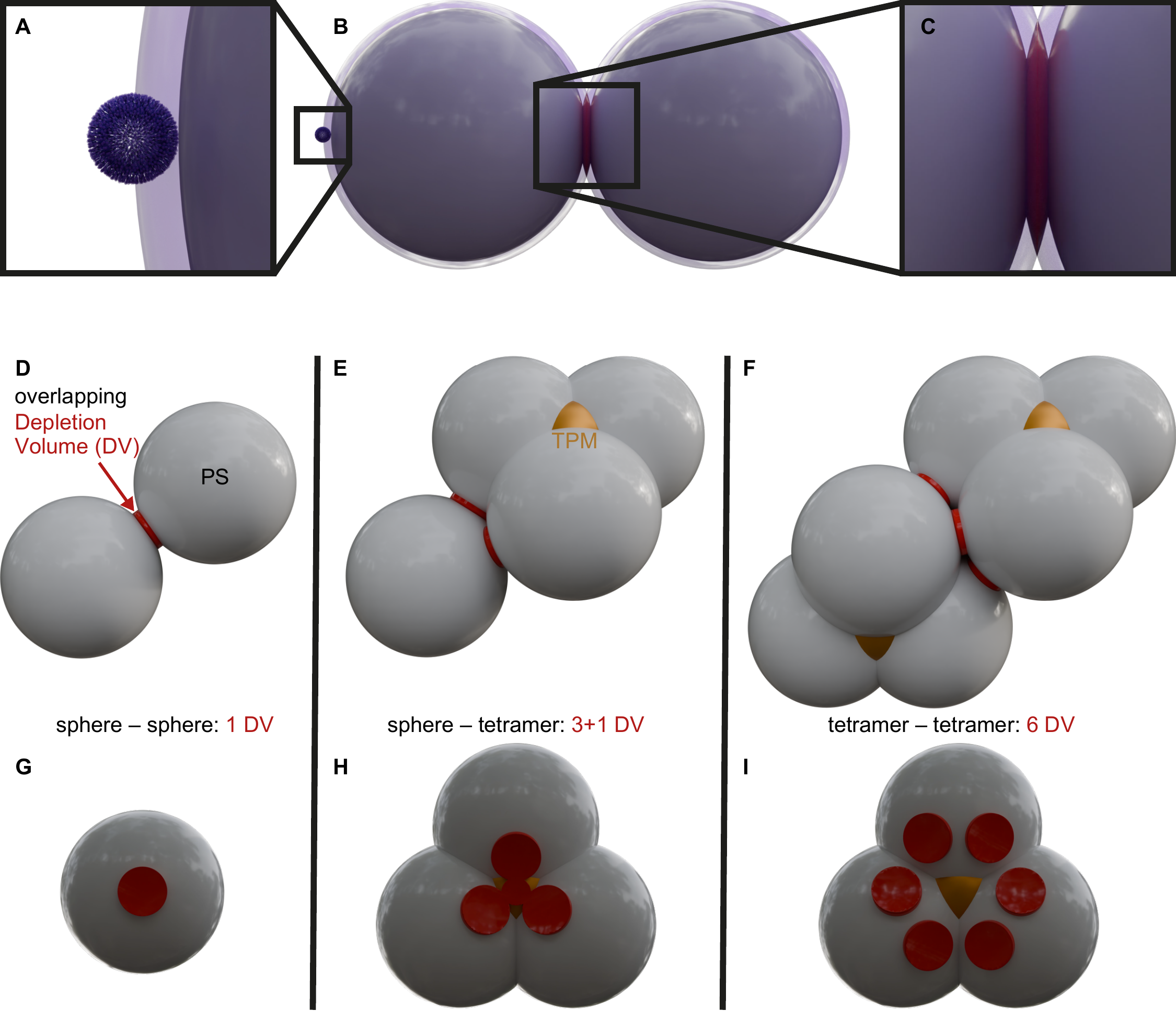}
    \caption{Overlapping depletion volumes: (A) A $\SI{15}{nm}$ micelle and the excluded volume around a $\SI{267}{nm}$ polystyrene sphere (light purple). (B, C) Bringing two spheres into contact causes their excluded volumes to overlap (shown in red).  
    Images in (D-F) compare the overlapping depletion volumes (DVs) for sphere-sphere contact (D, G) with one DV; sphere-tetramer contact (E, H) with three or four DVs; and interlocked tetramer-tetramer contact (F, I) with six DVs. In (E, H), the level of TPM extrusion determines the DV overlap (three or four).
    }
    \label{fig:DepletionVolume}
\end{figure}

However, an effective sphere description neglects cluster geometry, which becomes central for partially fused objects. Thus, a more detailed qualitative model that accounts for cluster geometry and is based on depletion-volume overlap is required (Fig.\,\ref{fig:DepletionVolume}). In the two interlocked tetramers shown in Fig.\,\ref{fig:DepletionVolume}F, I, three spheres of one tetramer touch three spheres of the adjacent one, giving rise to six contact points. In this idealised arrangement, the resulting excluded volume is 6 times that of isolated spheres in contact with each other (Fig.\,\ref{fig:DepletionVolume}D, G) and 3.8 times that of the effective sphere introduced above. To separate tetramers from uncoordinated spheres, the relevant intermediate case is a single sphere touching three tetramer spheres, as shown in Fig.\,\ref{fig:DepletionVolume}E, H. This results in an excluded volume that is 3 times the sum of the volumes of two spheres. Additionally, a sufficient extrusion of the TPM patch may lead to a further overlap of exclusion volumes (Fig.\,\ref{fig:DepletionVolume}H). Together, these considerations define an entropic hierarchy of three possible combinations:  (i) sphere-sphere, (ii) sphere-tetramer, and (iii) tetramer-tetramer. In this hierarchy, the excluded volumes and, consequently, the entropic driving force increase by a factor of three. Indeed, single spheres (red) are clustered with aggregates in Fig.\,\ref{fig:Tsort}A, B, indicating that they are predominantly trapped in the intermediate hierarchy level (ii).

The results of Fig.\,\ref{fig:Tsort} demonstrate that tetramers can be efficiently separated from uncoordinated spheres. Fig. \ref{fig:Tsort}D shows the separation of partially fused tetramers, and Fig.\,S5 shows the most extreme case of fully fused tetramers. Importantly, fusion systematically alters the geometry and reduces the accessible overlap volume. As tetramers deform towards a more spherical shape, the effective excluded volume decreases from a factor of six to $\sqrt[3]{4}$ (i.e.,\ by a factor of 3.8) compared to that of uncoordinated colloids. This effect can clearly be seen in Fig. \ref{fig:Tsort}D.

Although the geometric argument above predicts strong excluded volume interactions for colloidal clusters, their sorting efficiency appears to be lower than that shown in Fig.\,\ref{fig:sorting} for binary sphere mixtures, where nearly monodisperse phases can be achieved. This suggests that cluster phase separation may have slower kinetics than binary colloid blends. In particular, the probability that two approaching tetramers collide in the high-overlap “face-to-face” configuration (i.e., in the way shown in Fig.\,\ref{fig:DepletionVolume}F) is low. More commonly, tetramers will first contact via their single lobes, resulting in excluded volume interactions comparable to those of uncoordinated spheres. This implies that a larger number of collisions is required to reach strongly binding configurations, thereby slowing aggregation and, consequently, phase separation. Longer segregation times (or protocols that suppress gravitational sedimentation) may therefore be needed to achieve complete demixing in cluster suspensions. 

Finally, we revisit the thermodynamic argument described in the introduction, in which aggregation increases the entropy of smaller, more numerous micelles by releasing overlap volume. While tetramer segregation and subsequent crystallisation can release up to six times more overlap volume than single spheres (in the idealised interlocked configuration presented above), the entropy balance is more nuanced for non-spherical objects. On the one hand, non-spherical colloidal objects possess additional rotational degrees of freedom that are partially frozen out upon sedimentation. This alters the entropy balance in cluster segregation compared to that for perfect spheres. On the other hand, rapid rotational motion increases the effective volume occupied by tetramers in suspension, which can be approximated by the volume of a sphere around their contours. This rotationally averaged volume is larger by a factor of $(1+\sqrt{6}/2)^3/4\approx 2.75$ compared to unfused tetramers and by a factor of $\approx 1.73$ compared to the fully fused, spherical tetramer. This larger released volume should enhance tetramer segregation compared to the geometrical argument above. These counteracting contributions are not included in the simplified excluded-volume considerations above, motivating the development of a more complete model for depletion-mediated sorting of quantised cluster geometries (See Supplementary Material for a calculation of the depletion volumes). 

\section{Conclusions}
\noindent We demonstrate that excluded-volume interactions provide a practical approach to size-selective sorting of colloidal mixtures and can be applied to non-spherical colloidal clusters. By constructing stability phase diagrams for polystyrene spheres across three micelle-forming surfactants, we reveal that the mixing-demixing boundary occurs within a finite window of surfactant concentration, rather than at a sharp threshold. This transition width is particularly important at low size contrast. Instead of requiring two fully pure coexisting phases, this transition width can be exploited to obtain a single highly purified, nearly monodisperse fraction (either the supernatant or the sediment) by selecting a surfactant concentration that lies within the transition window of one population but outside that of the other. Using this strategy, we extend the sorting of binary sphere mixtures to size ratios of around 1.6 -- substantially below the values typically targeted in depletion-based fractionation. A comparably fine size discrimination has been achieved independently and concurrently via a sequential depletion-tuned crystallisation-melting protocol, which separates colloids differing in size by less than 10\% \cite{Yang2026FractionalCrystallization}, reinforcing that the transition-width and repeated-fractionation strategies explored here and elsewhere are converging on similarly high-resolution, scalable routes to colloidal purification.

Building on this result, we applied the same approach to cluster-spheres mixtures and achieved the efficient enrichment of coordinated clusters. Tetramers are enhanced by around 15-fold in the sediment compared to uncoordinated spheres. While separating clusters from spheres is robust, sorting clusters with different coordination numbers (e.g.,\ tetramers versus trimers/pentamers) is considerably more challenging. Importantly, sorting efficiency depends not only on effective size but also on aggregate geometry. Weakly fused, lobe-bearing tetramers separate more efficiently than strongly fused, more spherical clusters, consistent with contact-dependent depletion-volume overlap.  More generally, these non-spherical colloids exhibit an interplay of concave and convex surface curvatures. Once symmetry is broken, the local exclusion and the available global volume are expected to scale differently. As illustrated in Fig.\,\ref{fig:DepletionVolume}, our geometric arguments suggest that this interplay governs the depletion interactions between such particles. A full quantitative analysis of the effects of this interplay near the demixing transition, where contributions to the thermodynamics beyond simple excluded volume become significant, is a topic for future study.

Our results also highlight key open questions regarding the depletion-driven sorting of clusters. Compared with binary sphere mixtures, the segregation of tetramers appears to be more strongly influenced by kinetic limitations. This is probably because `face-to-face' contact configurations with high excluded-volume overlap are rare, so many collisions are required before strongly bound geometries form. This suggests the need for protocols that suppress gravitational sedimentation (e.g. density matching or slow rotation), as these would enable more direct kinetic control and approach complete demixing. Finally, a predictive theoretical framework that combines geometry, rotation, and micelle thermodynamics would help identify conditions under which different cluster coordination states could be reliably separated. 

From a practical perspective, excluded-volume-driven sorting has clear advantages. Unlike density-gradient centrifugation, it is easily scalable and increases the yield of recoverable clusters. Repeating the segregation step should improve both yield and purity, providing a versatile purification route for colloidal building blocks used in hierarchical assembly.

\section{Experimental Section}
\sloppy{
\textit{Materials}: Styrene monomer ($\geq\!99 \%$), potassium persulfate (KPS), 3-(trimethoxysilyl) propyl methacrylate ($\geq\!98\%$, TPM), (3-glycidyloxypropyl) trimethoxysilane ($\geq\,98\%$), glycerol, sodium chloride (NaCl), tetrahydrofuran (THF), 2,2'-Azobis-(2-methyl-propionitril) (AIBN), dodecyltrimethylammonium bromide (DTAB), sodium dodecylsulfate (SDS), Synperonic F108, Pluronic F127 and Triton X-100 were purchased from Sigma-Aldrich. Milli-Q water with a conductivity of 0.055\,\textmu S/cm was produced with an in-house purification system and used for all experiments.}

\textit{Polystyrene colloid synthesis}:
Polystyrene (PS) colloidal particles were synthesised using a well-established protocol via surfactant-free emulsion polymerisation \cite{PineNature2020}. A $\SI{1}{L}$ round-bottom flask immersed in an oil bath placed on a hotplate was used as a reactor. A cylindrical magnetic stirrer ($\SI{30}{mm} \times \SI{8}{mm}$) provided continuous vigorous stirring, and a nitrogen atmosphere with slight, constant purging was maintained. The ratio of radical initiator KPS to styrene monomer, which was previously filtered through an activated aluminium oxide column to remove any inhibitor, was kept constant at $\SI{0.01}{g}/\SI{1}{mL}$. The total amount of water was kept constant at $\SI{365}{mL}$. The amount of styrene monomer and reaction temperature were varied depending on the desired final size. Generally, smaller polystyrene colloidal spheres were produced at lower monomer concentration and higher reaction temperature. For large particles, a seeded growth technique was employed by replacing an equivalent volume of Milli-Q water in the reactor with a polystyrene colloid suspension before initiation. We use a seed size range of $\SI{200}{nm}$ to $\SI{550}{nm}$ to grow them from $\SI{470}{nm}$ up to $\SI{1068}{nm}$. We adjust the amount of seeds added relative to the styrene monomer and the total quantity for a specific seed size. We use monomer-to-seed ratios $m_\mathrm{monomer}/m_\mathrm{seed}$ of 5 to 10. This allows for tuning the growth ratio; generally, more relative seeds result in less monomer available per particle and a reduced growth ratio. By adjusting the overall amount, we can effectively suppress secondary growth.

\textit{Colloid imaging and particle sizing}:
After depletion-induced separation, the supernatant and precipitate were separately extracted. The washing procedure includes centrifugation using an Eppendorf Minispin Plus in $\SI{1.5}{mL}$ tubes at a relative centrifugal force $RCF_\mathrm{max}=7000$ for $\SI{10}{min}$, decantation, refilling with Milli-Q water, followed by redispersion using a vortex mixer. This cleaning procedure was repeated three times. In a final step after redispersion, the volume was increased to $\SI{100}{\micro\litre}$ by adding Milli-Q water.
For scanning electron microscopy (SEM), substrate $10\times 10\mathrm{mm^2}$ silicon wafers were first pre-cleaned via an Acetone ultrasonic bath for $\SI{10}{min}$, an acetone flush, an isopropanol rinse, and nitrogen dry-blowing. Surface hydrophilicity was increased by oxygen plasma using a Diener Zepto instrument working at $70\%$ power and 30\,mbar pressure for $\SI{5}{min}$. The wafers were mounted onto round aluminium SEM stubs ($\SI{12.5}{mm}$) with double-sided scotch tape. The top surface was subdivided into multiple regions using a permanent marker as hydrophobic barriers. Individual samples were drop-cast (about $\SI{20}{\micro\litre}$) onto silicon wafers and dried at room temperature. To minimise charging effects on the samples, a $\SI{4}{nm}$ thick conductive gold layer was sputtered onto them using a Cressington 208HR sputter coater. A copper band extending from the wafer top surface to the metal stub ensured a low-resistance conductive bridge. SEM imaging was performed using a Tescan Mira3 LM FE at an acceleration voltage of $3.5\mathrm{kV}$ (Fig.\,S5). ImageJ v1.54p \cite{Fiji2012} was used to determine the colloid diameters and to count colloids of different sizes in images.

Although dynamic light scattering (DLS) is a common method for analysing colloids, it was not employed in this study. The diameters of our colloids ranged from approximately 267 to 1008\,nm. In this size range, scattering increasingly falls within the Mie regime. This means that DLS-derived, intensity-weighted size distributions are sensitive to the optical model used, as well as to the presence of trace aggregates and the settling of larger particles. Consequently, DLS would not necessarily provide a more reliable estimate of the primary particle size distribution than direct scanning electron microscopy (SEM) analysis. Based on SEM measurements already performed, all particle batches exhibited coefficients of variation below 7\%, with most below 5\%, confirming their low polydispersity.

\textit{Colloidal cluster preparation}: 
Colloidal clusters were fabricated using a previously reported procedure \cite{PineNature2020}. First, monodisperse oil droplets containing 3-(trimethoxysilyl) propyl methacrylate and (3-glycidyloxypropyl) trimethoxysilane at a 4:1 volume ratio were prepared via self-emulsification under basic conditions ($\mathrm{pH}=10.8$) \cite{OilDropsLang2017}. The size of the resulting droplets was adjusted by adding Synperonic F108 surfactant ($1 \%\mathrm{w/v}$ solution), ranging from 10\,\textmu L to 1\,mL per 100\,mL of Milli-Q water. Second, after washing to remove the excess surfactant, the oil droplets were diluted to $\SI{40}{mL}$ added drop-by-drop to a large excess of polystyrene (PS) particles ($\SI{40}{mL}$ with $4 \%\mathrm{w/v}$ colloidal concentration, resulting in $N_\mathrm{Colloids}: N_\mathrm{Droplets}\geq 100: 1$) while both were kept at $\SI{50}{mM}$ sodium chloride to reduce charge stabilisation. This induced the formation of well-defined clusters, in which the size ratio between colloids and oil droplets determines how many colloids can be packed in contact with the droplet and, therefore, the coordination number. The suspension was stabilised by the addition of  1\,ml Synperionic F108 surfactant ($1 \%\mathrm{w/v}$ solution). After adding $\SI{47.2}{\%v/v}$ glycerol solution ($\SI{44}{\%w/w}$) with $\SI{52.8}{\%v/v}$ of the coordinated colloid suspension, isopycnic centrifugation was used to remove most of the uncoordinated PS particles from the resulting cluster suspension. Further sorting by coordination number was performed using the density-gradient and diffusion techniques. However, this approach has limitations in sorting quality for particle sizes below $\SI{1}{\um}$ and is low-throughput and time-consuming (a total processing time of 6 h for a single batch with a 3 mL, as above). Subsequently, after washing, the colloidal clusters were compressed and fused by adding controlled amounts of THF ($V_\mathrm{THF}/V_\mathrm{clusters}=0.5-1$) over varying periods of time ($\SI{30}{s}-\SI{120}{s}$) in the presence of $0.11\%\mathrm{w/v}$DTAB under vortex stirring. The shape of the oil patches extruding from the partially fused polystyrene spheres is determined by the interaction of the oil with the surfactant(s) in the quenching solution. We used typically $\SI{200}{\uL}$ of $1\%\mathrm{w/v}$ F127 and $\SI{50}{\uL}$ of $5\%\mathrm{w/v}$ SDS in $\SI{13}{mL}$ Milli-Q water. Finally, after washing, the soft oil patches and core were polymerised at $\SI{80}{^\circ C}$ for $\SI{2}{h}$ using AIBN as the radical initiator. The resulting hard clusters were then carefully washed.

\textit{Sample preparation}:
Stock solutions with varying concentrations of surfactants were prepared. They contain the three different surfactants Pluronic F-$127$ (Sigma-Aldrich) at $5, 10, 15, 20 \%\mathrm{w/v}$, sodium dodecyl sulfate (Sigma-Aldrich) at $5, 10, 15, 20 \%\mathrm{w/v}$, and Triton X-100 (Sigma-Aldrich) at $5, 10, 23.8 \%\mathrm{w/v}$ (Fig.\,\ref{fig:depletion}D). Higher concentrations increased the viscosity and caused the solutions to gel.
Stock solutions with varying salt concentrations of sodium chloride were prepared at $0.5, 5 \mathrm{M}$.

The weight concentrations of stock polystyrene particle suspensions were determined by comparing the weight of an empty $\SI{1.5}{mL}$ centrifuge vial with the weight of the vial filled with $\SI{1000}{\uL}$ of suspension before and after the complete water evaporation at $\SI{45}{^\circ C}$ for $\SI{72}{h}$. Taking an average of three measurements and using the polystyrene colloid density $\rho_\mathrm{PS, colloid}\!=\!\SI{1.050}{\frac{g}{cm^3}}$ \cite{minelli2018measuring}, we calculated the volume fraction and colloidal particle concentrations of all initial suspensions (from $1.75\%\mathrm{w/v}$ for the smallest $\SI{200}{nm}$ colloids to $9.7\%\mathrm{w/v}$ for the seeded grown ones). This allowed for the proper calculation of initial colloidal mixture concentrations and their comparison with those measured after fractionation. 

Samples of typically $\SI{2}{mL}$ were prepared by mixing calculated volumes of Milli-Q water, $\SI{5}{M}$ sodium chloride solution, particle stock suspension(s), and surfactant stock solution in $\SI{4}{mL}$ glass vials with screw caps. Pipette tips were primed for use with high-viscosity surfactant solutions. The reduced liquid volume due to the presence of the polystyrene core was accounted for when calculating salt and surfactant concentrations. The mixtures were homogenised using a vortex mixer at 2000\,rpm for $\SI{10}{s}$.

Closed vials were then placed on a low-vibration horizontal bench for $\SI{24}{h}$ to allow sample equilibration. To minimise the gravitational settling of the larger colloids ($d\!>\!\SI{1000}{nm}$), retrieval and evaluation of the samples was performed after $\SI{8}{h}$ or $\SI{4}{h}$. Alternatively, to separate the gravitational settling of individual colloids from the effect of excluded-volume-driven aggregation, the vials were flipped by 180\textdegree\ every $\SI{60}{min}$. Note that gravitational sedimentation is the same for all degrees of deformation because they originate from the same PS and TPM batches and have been exposed to slightly different concentrations of THF during the fabrication. Therefore, gravitational sedimentation reduces the quantitative advantage of shape-based sorting by depleting the population. However, the qualitative distinction of sorting by shape remains unaffected.

If only small amounts of sample were available, especially those involving coordinated particles, 0.5\,ml conical glass insert vials sealed with Parafilm were used instead. This ensured a proper extraction of the supernatant and precipitate from a sufficiently high column. 

\textit{Sample selection and retrieval}:
Once the settling and equilibration process due to the depletion effects had been completed, the regions of interest were selected. These regions were defined by their positions within the corresponding vertical column of the vial. A Hamilton syringe fitted with a flat-tip needle was manually inserted from the top to the selected height to remove the liquid from the layer and, if possible, all of the liquid initially present above it. To minimise turbulence and mixing by the needle, samples must be retrieved in top-down order. The supernatant was retrieved by directly removing the liquid phase containing the suspended colloids. The precipitate was obtained from the residue in the vial after all the liquid had been removed and redispersed.

\textit{Sample evaluation}:
The onset of excluded-volume-driven aggregation was determined for all colloid sizes and surfactant types as a function of surfactant concentration (Fig.\,\ref{fig:depletion} B, C). The amount of precipitation was visually estimated, with two limiting cases initially identified: the appearance of a precipitate phase at low surfactant concentrations and complete transparency of the supernatant phase at higher surfactant concentrations. Additional samples were then prepared with intermediate surfactant concentrations. The phase boundary was determined by subdivision of the samples into seven bins. Qualitatively, the height of the precipitation phase increased with each subsequent batch, while the opacity of the supernatant phase decreased. The two main observables are the turbidity of the supernatant and the amount of precipitate N: fully turbid with no precipitate; M0: fully turbid with some precipitate present; M1: fully turbid with a full layer of precipitate formed; M2: turbidity is reduced, with an observable interface between the supernatant and the precipitate; M3: medium turbidity with reduced whiteness; M4: slight turbidity, where the back of the vial is visible; Y: no turbidity, where only precipitate is present. This two-parameter classification allows for more precise distinctions than light transmittance measurements alone.

The phase boundary between the mixed and demixed phases was defined as the average of the two limiting cases, the difference between which was two times the standard deviation $\sigma$ (dashed and dotted lines in Fig.\,\ref{fig:depletion}c, respectively). To calculate the mean and weighted standard deviation of the phase transition, which serves as a measure of width, we assign a linear scale to these steps for weighting. The step weights are in the set $\in \{(\text{N},0), (\mathrm{M0}, 1), ..., (\text{Y},6)\}$. Using discrete numerical integration of a cumulative transition curve, we obtain the weighted mean as the first moment,  
\begin{equation}
    \mu =\frac{\sum_i\Delta w_i\cdot c_i^m}{\sum_i\Delta w_i}.
\end{equation}
Here, $c_i^m=\frac{1}{2}(c_i+c_{i+1})$ is the midpoint of two subsequent surfactant concentrations. The corresponding weights of the sample are given by $\Delta w_i = w_{i+1}-w_i$. Together with the second moment 
\begin{equation}
    \mu_2=\frac{\sum_i\Delta w_i\cdot (c_i^m)^2}{\sum_i\Delta w_i},
\end{equation}
we obtain the weighted standard deviation $\sigma=\sqrt{\mu_2-\mu^2}$. This can be seen by plotting the depletion against the Triton X-100 concentration for the 515\, nm-diameter polystyrene batch (Fig.\ref{fig:depletion}C, corresponding to the image in Fig.\ref{fig:depletion}B).
Repeating this experiment for all colloid sizes and surfactant types enabled us to construct the phase diagram shown in Fig.\,\ref{fig:depletion}E.

\medskip
\setlength\parindent{0pt}
\textbf{Supporting Information} \par
Supporting information is available online or from the author.

\medskip
\textbf{Acknowledgements} \par
Financial support from the Swiss National Science Foundation under grant number CRSII5-216659 is gratefully acknowledged. This project also benefited from funding through the National Competence Center in Bioinspired Materials (Grant No. 51NF40-182881) and the Adolphe Merkle Foundation. US acknowledges funding through the ERC Advanced Grant PrISMoID (Grant No.\ 833895). 

\medskip

\bibliographystyle{elsarticle-num}
\bibliography{literature}

@article{SANDERS:1964aa,
	author = {Sanders, J. V.},
	date = {1964/12/01},
	doi = {10.1038/2041151a0},
	id = {SANDERS1964},
	isbn = {1476-4687},
	journal = {Nature},
	number = {4964},
	pages = {1151--1153},
	title = {Colour of Precious Opal},
	url = {https://www.academia.edu/download/48107817/2041151a020160816-12727-1nyt4dm.pdf},
	volume = {204},
	year = {1964}}

@article{SteinerEmulsions1995,
	author = {Steiner, Ullrich and Meller, Amit and Stavans, Joel},
	doi = {10.1103/PhysRevLett.74.4750},
	issue = {23},
	journal = {Phys. Rev. Lett.},
	month = {Jun},
	numpages = {0},
	pages = {4750--4753},
	publisher = {American Physical Society},
	title = {Entropy Driven Phase Separation in Binary Emulsions},
	url = {https://link.aps.org/doi/10.1103/PhysRevLett.74.4750},
	volume = {74},
	year = {1995}}

@article{Bibette1991,
	author = {J. Bibette},
	doi = {https://doi.org/10.1016/0021-9797(91)90181-7},
	issn = {0021-9797},
	journal = {Journal of Colloid and Interface Science},
	number = {2},
	pages = {474-478},
	title = {Depletion interactions and fractionated crystallization for polydisperse emulsion purification},
	url = {https://www.sciencedirect.com/science/article/pii/0021979791901817},
	volume = {147},
	year = {1991}}

@article{OilDropsLang2017,
	author = {van der Wel, Casper and Bhan, Rohit K. and Verweij, Ruben W. and Frijters, Hans C. and Gong, Zhe and Hollingsworth, Andrew D. and Sacanna, Stefano and Kraft, Daniela J.},
	date = {2017/08/22},
	doi = {10.1021/acs.langmuir.7b01398},
	isbn = {0743-7463},
	journal = {Langmuir},
	journal1 = {Langmuir},
	journal2 = {Langmuir},
	month = {08},
	number = {33},
	pages = {8174--8180},
	publisher = {American Chemical Society},
	title = {Preparation of Colloidal Organosilica Spheres through Spontaneous Emulsification},
	type = {doi: 10.1021/acs.langmuir.7b01398},
	url = {https://doi.org/10.1021/acs.langmuir.7b01398},
	volume = {33},
	year = {2017},
	year1 = {2017}}

@article{PuseyNature1986,
	author = {Pusey, P. N. and van Megen, W.},
	date = {1986/03/01},
	doi = {10.1038/320340a0},
	id = {Pusey1986},
	isbn = {1476-4687},
	journal = {Nature},
	number = {6060},
	pages = {340--342},
	title = {Phase behaviour of concentrated suspensions of nearly hard colloidal spheres},
	url = {https://doi.org/10.1038/320340a0},
	volume = {320},
	year = {1986}}

@article{PineNature2020,
	author = {He, Mingxin and Gales, Johnathon P. and Ducrot, {\'E}tienne and Gong, Zhe and Yi, Gi-Ra and Sacanna, Stefano and Pine, David J.},
	date = {2020/09/01},
	doi = {10.1038/s41586-020-2718-6},
	id = {He2020},
	isbn = {1476-4687},
	journal = {Nature},
	number = {7826},
	pages = {524--529},
	title = {Colloidal diamond},
	url = {https://doi.org/10.1038/s41586-020-2718-6},
	volume = {585},
	year = {2020}}

@article{AsakuraOosawa54,
    author = {Asakura, Sho and Oosawa, Fumio},
    title = {On Interaction between Two Bodies Immersed in a Solution of Macromolecules},
    journal = {The Journal of Chemical Physics},
    volume = {22},
    number = {7},
    pages = {1255-1256},
    year = {1954},
    month = {07},
    issn = {0021-9606},
    doi = {10.1063/1.1740347},
    url = {https://doi.org/10.1063/1.1740347},
    eprint = {https://pubs.aip.org/aip/jcp/article-pdf/22/7/1255/18805169/1255_2_online.pdf},
}

@article{AsakuraOosawa58,
author = {Asakura, Sho and Oosawa, Fumio},
title = {Interaction between particles suspended in solutions of macromolecules},
journal = {Journal of Polymer Science},
volume = {33},
number = {126},
pages = {183-192},
doi = {https://doi.org/10.1002/pol.1958.1203312618},
url = {https://onlinelibrary.wiley.com/doi/abs/10.1002/pol.1958.1203312618},
eprint = {https://onlinelibrary.wiley.com/doi/pdf/10.1002/pol.1958.1203312618},
year = {1958}
}

@article{mirgorod2019structure,
  title={Structure of micelles of sodium dodecyl sulphate in water: An X-ray and dynamic light scattering study},
  author={Mirgorod, Yu and Chekadanov, Alexander and Dolenko, Tatiana},
  journal={Chemistry journal of Moldova},
  volume={14},
  number={1},
  pages={107--119},
  year={2019}
}

@incollection{stubivcar1989size,
  title={Size, shape and internal structure of Triton X-100 micelles determined by light and small-angle X-ray scattering techniques},
  author={Stubi{\v{c}}ar, N and Mateja{\v{s}}, J and Zipper, P and Wilfing, R},
  booktitle={Surfactants in solution},
  pages={181--195},
  year={1989},
  publisher={Springer}
}

@article{gentile2010thermogelation,
  title={Thermogelation analysis of F127-water mixtures by physical chemistry techniques},
  author={Gentile, Luigi and De Luca, Giuseppina and Antunes, Filipe E and Rossi, Cesare Oliviero and Ranieri, Giuseppe Antonio and others},
  journal={Appl. Rheol},
  volume={20},
  number={5},
  pages={52081--1},
  year={2010}
}

@article{Fiji2012,
  title        = {Fiji: an open-source platform for biological-image analysis},
  author       = {Schindelin, Johannes and Arganda-Carreras, Ignacio and Frise, Erwin and Kaynig, Verena and Longair, Mark and Pietzsch, Tobias and Preibisch, Stephan and Rueden, Curtis and Saalfeld, Stephan and Schmid, Benjamin and Tinevez, Jean-Yves and White, Daniel J. and Hartenstein, Volker and Eliceiri, Kevin and Tomancak, Pavel and Cardona, Albert},
  journal      = {Nature Methods},
  volume       = {9},
  number       = {7},
  pages        = {676--682},
  year         = {2012},
  publisher    = {Nature Publishing Group}
}

@article{minelli2018measuring,
	author = {Minelli, Caterina and Sikora, Aneta and Garcia-Diez, Raul and Sparnacci, Katia and Gollwitzer, Christian and Krumrey, Michael and Shard, Alex G.},
	doi = {10.1039/C8AY00237A},
	issue = {15},
	journal = {Anal. Methods},
	pages = {1725-1732},
	publisher = {The Royal Society of Chemistry},
	title = {Measuring the size and density of nanoparticles by centrifugal sedimentation and flotation},
	url = {http://dx.doi.org/10.1039/C8AY00237A},
	volume = {10},
	year = {2018}}

@article{Glotzer2007Anisotropy,
  author  = {Glotzer, S. C. and Solomon, M. J.},
  title   = {Anisotropy of building blocks and their assembly into complex structures},
  journal = {Nature Materials},
  volume  = {6},
  pages   = {557--562},
  year    = {2007}
}

@article{Sacanna2011Nature,
  author  = {Sacanna, S. and Irvine, W. T. M. and Chaikin, P. M. and Pine, D. J.},
  title   = {Lock and key colloids},
  journal = {Nature},
  volume  = {464},
  pages   = {575--578},
  year    = {2010}
}

@book{Israelachvili2011,
  author    = {Israelachvili, J. N.},
  title     = {Intermolecular and Surface Forces},
  edition   = {3},
  publisher = {Academic Press},
  year      = {2011}
}

@article{Vrij1976Depletion,
  author  = {Vrij, A.},
  title   = {Polymers at interfaces and the interactions in colloidal dispersions},
  journal = {Pure and Applied Chemistry},
  volume  = {48},
  pages   = {471--483},
  year    = {1976}
}

@book{Lekkerkerker2011Colloids,
  author    = {Lekkerkerker, H. N. W. and Tuinier, R.},
  title     = {Colloids and the Depletion Interaction},
  publisher = {Springer},
  year      = {2011}
}

@book{Rosen2012Surfactants,
  author    = {Rosen, M. J. and Kunjappu, J. T.},
  title     = {Surfactants and Interfacial Phenomena},
  edition   = {4},
  publisher = {Wiley},
  year      = {2012}
}

@book{Hunter2001Zeta,
  author    = {Hunter, R. J.},
  title     = {Foundations of Colloid Science},
  publisher = {Oxford University Press},
  year      = {2001}
}

@article{Poon2002Phase,
  author  = {Poon, W. C. K.},
  title   = {The physics of a model colloid--polymer mixture},
  journal = {Journal of Physics: Condensed Matter},
  volume  = {14},
  pages   = {R859--R880},
  year    = {2002}
}

@article{Sacanna2013Shaping,
  author  = {Sacanna, S. and Pine, D. J.},
  title   = {Shape-anisotropic colloids: Building blocks for complex assemblies},
  journal = {Current Opinion in Colloid \& Interface Science},
  volume  = {18},
  pages   = {484--488},
  year    = {2013}
}

@article{Chen2011Fusion,
  author  = {Chen, Q. and Bae, S. C. and Granick, S.},
  title   = {Directed self-assembly of a colloidal kagome lattice},
  journal = {Nature},
  volume  = {469},
  pages   = {381--384},
  year    = {2011}
}

@article{onsager1949,
  title={The effects of shape on the interaction of colloidal particles},
  author={Onsager, Lars},
  journal={Annals of the New York Academy of Sciences},
  volume={51},
  number={4},
  pages={627--659},
  year={1949},
  doi={10.1111/j.1749-6632.1949.tb27296.x}
}

@book{doi1986,
  title={The Theory of Polymer Dynamics},
  author={Doi, Masao and Edwards, Samuel F.},
  year={1986},
  publisher={Oxford University Press},
  address={Oxford}
}

@article{frenkel1988,
  title={Entropy-driven phase transitions},
  author={Frenkel, Daan},
  journal={Physica A},
  volume={154},
  number={3},
  pages={424--440},
  year={1988},
  doi={10.1016/0378-4371(88)90066-7}
}

@article{poon2004,
  title={Colloids as big atoms},
  author={Poon, Wilson C. K.},
  journal={Science},
  volume={304},
  pages={830--831},
  year={2004},
  doi={10.1126/science.1096549}
}

@book{verwey1948,
  title={Theory of the Stability of Lyophobic Colloids},
  author={Verwey, E. J. W. and Overbeek, J. Th. G.},
  publisher={Elsevier},
  year={1948}
}

@article{frenkel1999,
  title={Entropy-driven ordering in colloids},
  author={Frenkel, Daan},
  journal={Physica A},
  volume={263},
  pages={26--38},
  year={1999},
  doi={10.1016/S0378-4371(98)00487-5}
}

@Article{C5SM02038G,
author ="Kang, Louis and Gibaud, Thomas and Dogic, Zvonimir and Lubensky, T. C.",
title  ="Entropic forces stabilize diverse emergent structures in colloidal membranes",
journal  ="Soft Matter",
year  ="2016",
volume  ="12",
issue  ="2",
pages  ="386-401",
publisher  ="The Royal Society of Chemistry",
doi  ="10.1039/C5SM02038G",
url  ="http://dx.doi.org/10.1039/C5SM02038G"}

@Article{C8SM02048E,
author ="Ballard, Nicholas and Law, Adam D. and Bon, Stefan A. F.",
title  ="Colloidal particles at fluid interfaces: behaviour of isolated particles",
journal  ="Soft Matter",
year  ="2019",
volume  ="15",
issue  ="6",
pages  ="1186-1199",
publisher  ="The Royal Society of Chemistry",
doi  ="10.1039/C8SM02048E",
url  ="http://dx.doi.org/10.1039/C8SM02048E"}

@article{AVENDANO201762,
title = {Packing, entropic patchiness, and self-assembly of non-convex colloidal particles: A simulation perspective},
journal = {Current Opinion in Colloid \& Interface Science},
volume = {30},
pages = {62-69},
year = {2017},
issn = {1359-0294},
doi = {https://doi.org/10.1016/j.cocis.2017.05.005},
url = {https://www.sciencedirect.com/science/article/pii/S1359029417300493},
author = {Carlos Avendaño and Fernando A. Escobedo}}

@article{
doi:10.1126/science.1253751,
author = {Vinothan N. Manoharan },
title = {Colloidal matter: Packing, geometry, and entropy},
journal = {Science},
volume = {349},
number = {6251},
pages = {1253751},
year = {2015},
doi = {10.1126/science.1253751},
URL = {https://www.science.org/doi/abs/10.1126/science.1253751},
eprint = {https://www.science.org/doi/pdf/10.1126/science.1253751}}

@article{XING201554,
title = {Depletion versus stabilization induced by polymers and nanoparticles: The state of the art},
journal = {Current Opinion in Colloid \& Interface Science},
volume = {20},
number = {1},
pages = {54-59},
year = {2015},
issn = {1359-0294},
doi = {https://doi.org/10.1016/j.cocis.2014.11.012},
url = {https://www.sciencedirect.com/science/article/pii/S1359029414001459},
author = {Xiaochen Xing and Li Hua and To Ngai}
}

@article{BRISCOE201546,
title = {Depletion forces between particles immersed in nanofluids},
journal = {Current Opinion in Colloid \& Interface Science},
volume = {20},
number = {1},
pages = {46-53},
year = {2015},
issn = {1359-0294},
doi = {https://doi.org/10.1016/j.cocis.2014.12.002},
url = {https://www.sciencedirect.com/science/article/pii/S1359029414001472},
author = {Wuge H. Briscoe}
}

@article{10.1063/5.0173340,
    author = {He, Yukun and Qiao, Yi and Ding, Lu and Cheng, Tianguang and Tu, Jing},
    title = {Recent advances in droplet sequential monitoring methods for droplet sorting},
    journal = {Biomicrofluidics},
    volume = {17},
    number = {6},
    pages = {061501},
    year = {2023},
    month = {11},
    issn = {1932-1058},
    doi = {10.1063/5.0173340},
    url = {https://doi.org/10.1063/5.0173340},
    eprint = {https://pubs.aip.org/aip/bmf/article-pdf/doi/10.1063/5.0173340/18208129/061501_1_5.0173340.pdf},
}

@article{https://doi.org/10.1002/pi.4842,
author = {Robin, Mathew P and O'Reilly, Rachel K},
title = {Strategies for preparing fluorescently labelled polymer nanoparticles},
journal = {Polymer International},
volume = {64},
number = {2},
pages = {174-182},
doi = {https://doi.org/10.1002/pi.4842},
url = {https://scijournals.onlinelibrary.wiley.com/doi/abs/10.1002/pi.4842},
eprint = {https://scijournals.onlinelibrary.wiley.com/doi/pdf/10.1002/pi.4842},
year = {2015}
}

@article{https://doi.org/10.1002/anie.201001451,
author = {Li, Fan and Josephson, David P. and Stein, Andreas},
title = {Colloidal Assembly: The Road from Particles to Colloidal Molecules and Crystals},
journal = {Angewandte Chemie International Edition},
volume = {50},
number = {2},
pages = {360-388},
doi = {https://doi.org/10.1002/anie.201001451},
url = {https://onlinelibrary.wiley.com/doi/abs/10.1002/anie.201001451},
eprint = {https://onlinelibrary.wiley.com/doi/pdf/10.1002/anie.201001451},
year = {2011}
}

@article{PhysRevLett.99.268301,
  title = {Directing Colloidal Self-Assembly through Roughness-Controlled Depletion Attractions},
  author = {Zhao, Kun and Mason, Thomas G.},
  journal = {Phys. Rev. Lett.},
  volume = {99},
  issue = {26},
  pages = {268301},
  numpages = {4},
  year = {2007},
  month = {Dec},
  publisher = {American Physical Society},
  doi = {10.1103/PhysRevLett.99.268301},
  url = {https://link.aps.org/doi/10.1103/PhysRevLett.99.268301}
}

@article{Neophytou2021ColloidalDiamond,
	author = {Neophytou, Andreas and Chakrabarti, Dwaipayan and Sciortino, Francesco},
	doi = {10.1073/pnas.2109776118},
	journal = {Proceedings of the National Academy of Sciences},
	number = {48},
	pages = {e2109776118},
	title = {Facile self-assembly of colloidal diamond from tetrahedral patchy particles via ring selection},
	volume = {118},
	year = {2021}
    }

@article{Baran2023PursuingColloidalDiamonds,
	author = {Baran, {\L}ukasz and Tarasewicz, Dariusz and Kami{\'n}ski, Daniel M. and R{\.z}ysko, Wojciech},
	doi = {10.1039/D3NR01771K},
	journal = {Nanoscale},
	pages = {10623--10633},
	title = {Pursuing colloidal diamonds},
	volume = {15},
	year = {2023}
    }

@article{Kim2023PatchyColloidalClusters,
	author = {Kim, You-Jin and Moon, Jeong-Bin and Hwang, Hyerim and Kim, Youn Soo and Yi, Gi-Ra},
	doi = {10.1002/adma.202203045},
	journal = {Advanced Materials},
	number = {4},
	pages = {2203045},
	title = {Advances in Colloidal Building Blocks: Toward Patchy Colloidal Clusters},
	volume = {35},
	year = {2023}
    }

@article{vanKesteren2024FACS,
	author = {van Kesteren, Steven and Diethelm, Pascal and Isa, Lucio},
	doi = {10.1039/D4SM00122B},
	journal = {Soft Matter},
	number = {13},
	pages = {2881--2886},
	title = {Fluorescence-activated cell sorting ({FACS}) for purifying colloidal clusters},
	volume = {20},
	year = {2024}
    }

@book{Lekkerkerker2024Depletion,
	address = {Cham},
	author = {Lekkerkerker, Henk N. W. and Tuinier, Remco and Vis, Mark},
	doi = {10.1007/978-3-031-52131-7},
	edition = {2nd},
	publisher = {Springer},
	series = {Lecture Notes in Physics},
	title = {Colloids and the Depletion Interaction},
	volume = {1026},
	year = {2024}
    }

@article{Ofosu2025NanocrystalDepletion,
	author = {Ofosu, Charles K. and Wilcoxson, Tanner A. and Lee, Tsung-Lun and Brackett, William D. and Choi, Jinny and Truskett, Thomas M. and Milliron, Delia J.},
	doi = {10.1126/sciadv.adv2216},
	journal = {Science Advances},
	number = {15},
	pages = {eadv2216},
	title = {Assessing depletion attractions between colloidal nanocrystals},
	volume = {11},
	year = {2025}
    }

@article{Tong2026SurfaceDictatedSorting,
	author = {Tong, Jintao and Zang, Shihao and Dou, Xiangyu and Chen, Xingyue and Chen, Chen and Xu, Zhe and Ma, Cheng and Huang, Jianbin},
	doi = {10.1016/j.jcis.2026.140421},
	journal = {Journal of Colloid and Interface Science},
	pages = {140421},
	title = {Surface-dictated sorting in colloidal mixtures},
	volume = {717},
	year = {2026}
    }

@article{Yang2026FractionalCrystallization,
	author = {Yang, Sehee and Park, Sanghyuk and Jung, Yongseok and Kim, Young Geon and Lee, Jiwoo and Kim, Shin-Hyun},
	doi = {10.1038/s41467-026-72000-y},
	journal = {Nature Communications},
	pages = {2026},
	title = {Fractional crystallization of colloids by sequential tuning of depletion forces},
	volume = {17},
	year = {2026}
    }
\includepdf[pages=-]{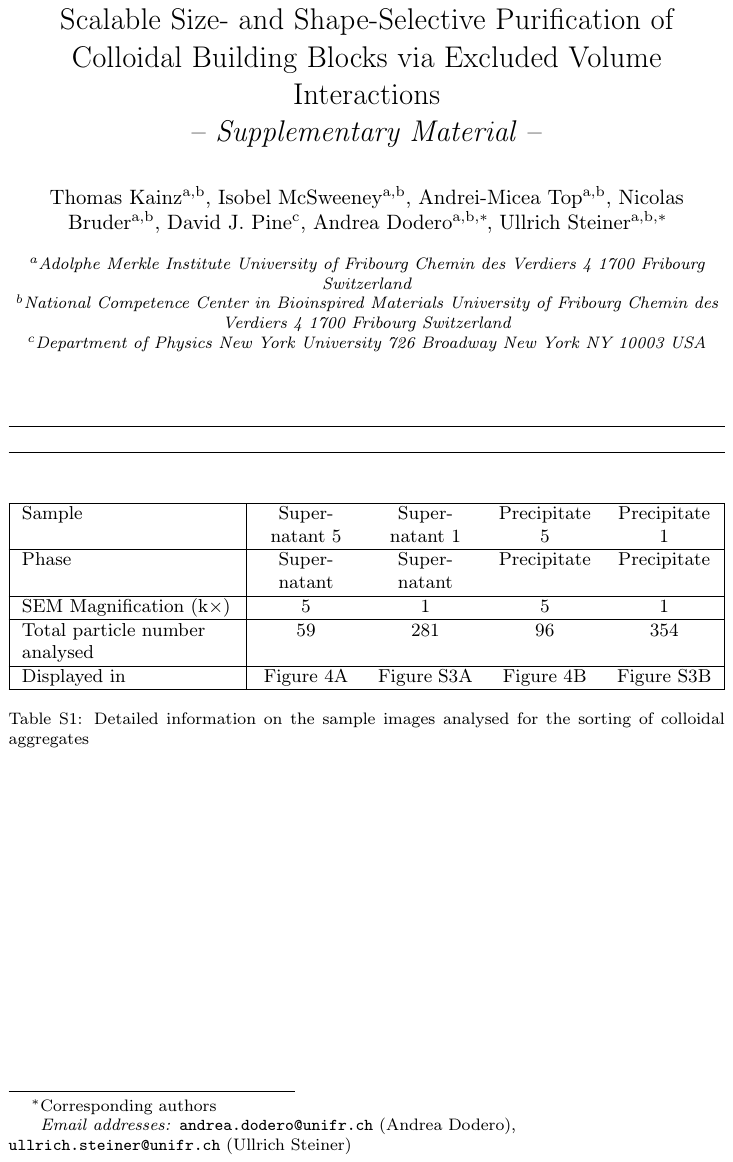}
\end{document}